\documentclass[prl,twocolumn,superscriptaddress,nobibnotes,a4paper]{revtex4-2}
\usepackage[pdftex]{graphicx}
\usepackage[pdftex]{epsfig}
\usepackage{amsmath}
\usepackage{amssymb}
\usepackage{amsfonts}
\usepackage{color}
\usepackage{wrapfig}
\usepackage{eucal}
\usepackage{hhline}
\usepackage{threeparttable}
\usepackage{supertabular}
\usepackage{multirow}
\usepackage{tabularx}
\usepackage{siunitx}
\usepackage{upgreek}
\usepackage{float}

\newcommand\numberthis{\addtocounter{equation}{1}\tag{\theequation}}

\usepackage[export]{adjustbox}

\usepackage{soul}

\usepackage[version=4]{mhchem} 
\usepackage{array} 
\usepackage[colorlinks=true, allcolors=blue]{hyperref} 

\newcommand{\ket}[1]{\left|#1\right\rangle}

\newcommand{\phase}[1]{\phi_\text{{#1}}}
\newcommand{\pphase}[2]{\phi_\text{{#1}}^{#2}}

\setstcolor{red}

\begin{document}

\title{On-chip distribution of quantum information using traveling phonons}\thanks{This work was published in \href{https://doi.org/10.1126/sciadv.add2811}{Science Adv.\ \textbf{8}, eadd2811 (2022).}}

\author{Amirparsa Zivari}\thanks{These authors contributed equally to this work.}
\affiliation{Kavli Institute of Nanoscience, Department of Quantum Nanoscience, Delft University of Technology, 2628CJ Delft, The Netherlands}
\author{Niccol\`{o} Fiaschi}\thanks{These authors contributed equally to this work.}
\affiliation{Kavli Institute of Nanoscience, Department of Quantum Nanoscience, Delft University of Technology, 2628CJ Delft, The Netherlands}
\author{Roel Burgwal}
\affiliation{Center for Nanophotonics, AMOLF, Science Park 104, 1098XG Amsterdam, The Netherlands}
\affiliation{Department of Applied Physics, Eindhoven University of Technology, P.O.\ Box 513, 5600MB Eindhoven, The Netherlands}
\author{Ewold Verhagen}
\affiliation{Center for Nanophotonics, AMOLF, Science Park 104, 1098XG Amsterdam, The Netherlands}
\affiliation{Department of Applied Physics, Eindhoven University of Technology, P.O.\ Box 513, 5600MB Eindhoven, The Netherlands}
\author{Robert Stockill}
\affiliation{Kavli Institute of Nanoscience, Department of Quantum Nanoscience, Delft University of Technology, 2628CJ Delft, The Netherlands}
\author{Simon Gr\"oblacher}
\email{s.groeblacher@tudelft.nl}
\affiliation{Kavli Institute of Nanoscience, Department of Quantum Nanoscience, Delft University of Technology, 2628CJ Delft, The Netherlands}


\begin{abstract}
Distributing quantum entanglement on a chip is a crucial step towards realizing scalable quantum processors. Using traveling phonons -- quantized guided mechanical wavepackets -- as a medium to transmit quantum states is currently gaining significant attention, due to their small size and low propagation speed compared to other carriers, such as electrons or photons. Moreover, phonons are highly promising candidates to connect heterogeneous quantum systems on a chip, such as microwave and optical photons for long-distance transmission of quantum states via optical fibers. Here, we experimentally demonstrate the feasibility of distributing quantum information using phonons, by realizing quantum entanglement between two traveling phonons and creating a time-bin encoded traveling phononic qubit. The mechanical quantum state is generated in an optomechanical cavity and then launched into a phononic waveguide in which it propagates for around 200 micrometers. We further show how the phononic, together with a photonic qubit, can be used to violate a Bell-type inequality.	
\end{abstract}


\maketitle

\section{Introduction}

Over the past decades, quantum technologies have evolved from scientific proof-of-principle experiments to a nascent and thriving industry. With recent demonstrations of quantum advantage over classical computation in multiple systems~\cite{Arute2019,Zhong2020}, the need for connecting such resources is becoming ever more urgent. Distributing quantum entanglement between distant parties is a crucial step towards implementing quantum repeaters and networks~\cite{Kimble2008,Sangouard2011}. Distributing it on-chip is needed for sparse qubit array architectures, which require on-chip long-range qubit couplers~\cite{Vandersypen2017}. Additionally, having entanglement between a stationary quantum memory and a flying qubit plays a central role in low-loss quantum information transfer over long distances~\cite{Kimble2008}.

One of the key challenges of building a quantum network is forming interfaces between heterogeneous quantum devices. A highly versatile system for this task has been identified in phonons, which can act as efficient intermediaries between different resources~\cite{Wallquist2009}. In particular, phonons have been shown to be highly useful in converting states between different optical wavelengths~\cite{Hill2012}, as well as for microwave to optics frequency conversion~\cite{Vainsencher2016,Andrews2014,Han2020,Arnold2020,Jiang2020,Forsch2020,Stockill2022}. Most recently, such a mechanical transducer has been used to transfer signals from a superconducting qubit to an optical fiber~\cite{Mirhosseini2020}, a key step for quantum information transfer. Moreover, the potential for quantum gate operations using phonons has been shown~\cite{Palomaki2013,Chu2017,Blatt2008}, owing to long coherence times and high transfer fidelities of phonons. The interest in traveling phonons in fact goes well beyond enabling long-distance quantum networks. Several of the most exciting prospects are arising from their many orders of magnitude slower propagation speed compared to light, low loss transmission and their small mode volume compared to traveling GHz photons. These unique features could have the potential to enable the on-chip distribution and processing of quantum information in a highly compact fashion~\cite{Delsing2019}, allow for coherent interactions with a large variety of quantum systems such as defect centers~\cite{Golter2016}, superconducting qubits~\cite{Bienfait2019} and quantum dots~\cite{McNeil2011,Hermelin2011}, in both homogeneous or heterogeneous implementations~\cite{Neuman2021}. Demonstrating the basic building blocks, such as marking the distribution of quantum information using highly confined phonons, remains an open challenge to date.

\begin{figure*}[ht!]
	\includegraphics[width = 0.9 \linewidth]{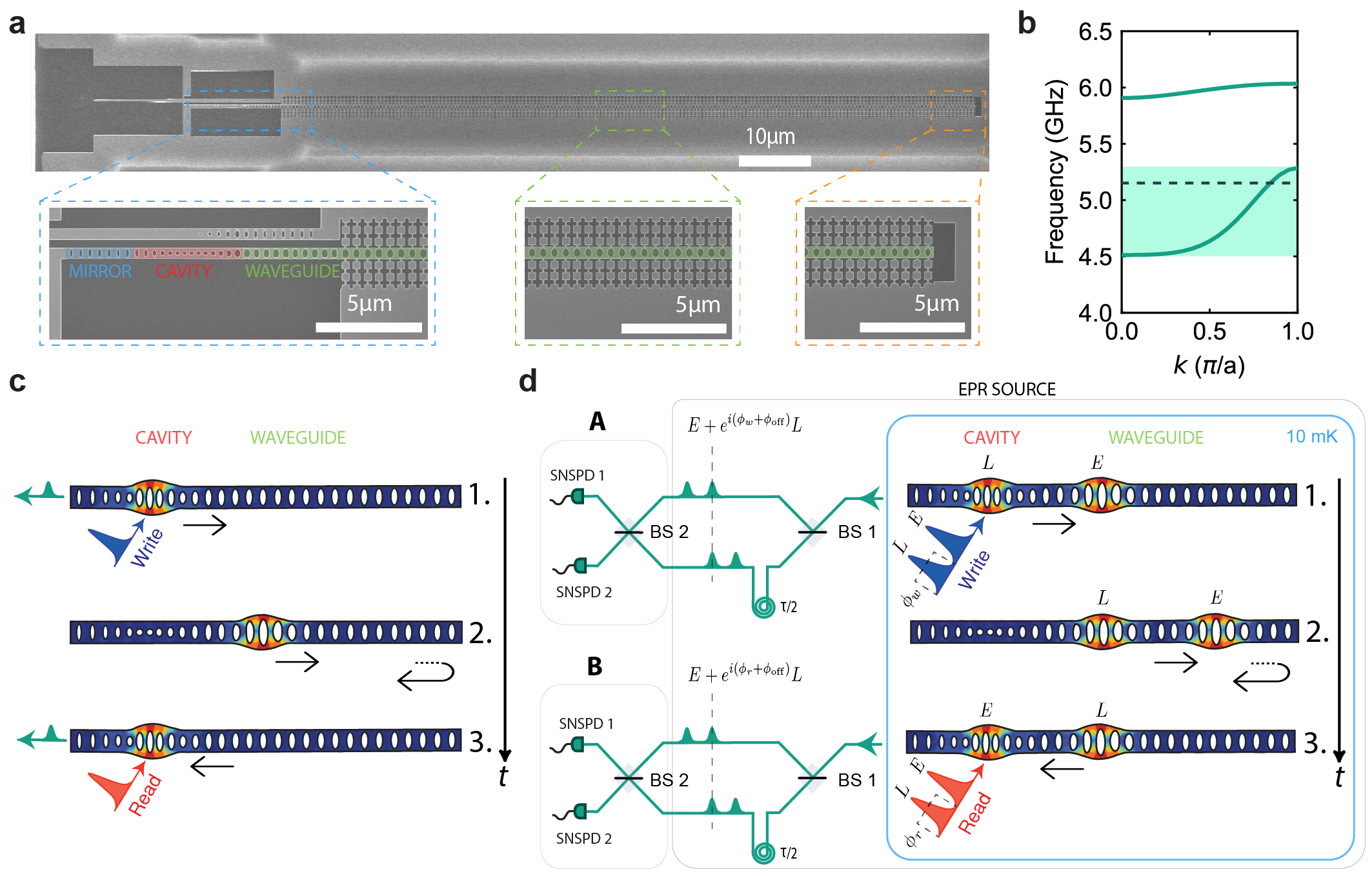}
	\caption{\textbf{Device and experimental setup.} a) Scanning electron microscope (SEM) image of the device. Bottom left:\ photonic and phononic mirror (highlighted in blue), optomechanical cavity (red) and initial part of the phononic waveguide, which also acts as a photonic mirror (green). Bottom center:\ section of the phononic waveguide. Bottom right:\ free standing end of the waveguide, which acts as a mirror for the phonons. b) Band diagram of a unit cell of the waveguide showing its single mode design for the symmetric mode, with the frequency of interest depicted by the black dashed line (see the SI section 1). c) The control pulses (write and read) are sent to the cavity to create 1.\ and retrieve 3.\ the mechanical excitation. The green pulses depict the scattered Stokes and anti-Stokes photons. In 2.\ the mechanical excitation travels in the waveguide (round-trip time of $\tau$). d) Simplified schematics of the time-bin entangling protocol. 1.\ Creation of the entangled state between the Stokes-scattered photons and the traveling phononic excitation in the waveguide. The pulses have a time delay of $\tau/2$ and are depicted here in shorter succession than in the experiment for clarity of the drawing. 2.\ Propagation of the mechanical qubit in the waveguide, with the reflection at the end. 3.\ Mapping of the phononic onto a photonic state in order to verify the entanglement. The boxes conceptually divide Einstein–Podolsky–Rosen (EPR) source, and measurements A and B (which are the same experimental setup at different times $t$), used to create and detect the entangled state. SNSPDs are superconducting nanowire single-photon detectors. The arrows in c) and d) represent the direction of propagation of the mechanical excitations and the time axis is shown by the vertical black arrow.}
	\label{Fig:1}
\end{figure*}

Quantum optomechanics has proven to be a versatile toolbox for controlling stationary, strongly confined phonons~\cite{Aspelmeyer2014,Barzanjeh2022}. Previously, bulk and surface acoustic waves (BAWs and SAWs, respectively) have been shown to be able to operate in the quantum regime~\cite{Chu2017,Delsing2019}, for example, by coupling to superconducting qubits for transducer and quantum information applications~\cite{Ekstrom2017,Dumur2021}, as well as entangling acoustic phonons~\cite{Bienfait2020}. These systems benefit from deterministic quantum operations with high fidelities, enabled by the non-linearity of the superconducting qubit and strong coupling between the qubit and the phononic channel. However, the confinement of the phonons in optomechanical devices results in several advantages, such as stronger field coupling~\cite{Chan2012}, higher coherence and longer lifetime~\cite{Wallucks2020}, and long distance routing capability on chip~\cite{Patel2018,Zivari2022}. In this work, building on these recent developments, we experimentally distribute quantum information using phonons in a waveguide. Our device is composed of an optomechanical cavity, which acts as the single phonon source and detector, that is connected to a single-mode phononic waveguide. Using this device we then create a time-bin entangled state of a pair of traveling phonons. Furthermore, we unambiguously show the non-classical correlations between an optical and the traveling phononic qubit, by violating a Bell-type inequality~\cite{Clauser1969,Marinkovic2018,Velez2020}.

\section{Results}
Our device consists of a single mode optomechanical cavity connected to a phononic waveguide (cf.\ Fig.~\ref{Fig:1}a), similar to a previous design~\cite{Zivari2022}. The cavity is used as a source and detector for mechanical excitations, controlled with telecom-wavelength optical pulses (via Stokes and anti-Stokes scattering~\cite{Riedinger2016}). We engineer the photonic and phononic band structure of the different parts of the device such that the mechanical mode extends into the waveguide while the optical mode remains fully confined in the cavity (see the Supplementary Information (SI) for more details on the device design). The waveguide has a free standing end, which acts as mirror for the traveling phonons and effectively forms a Fabry-P\'{e}rot cavity. The coupling between the single mode cavity and this Fabry-P\'{e}rot cavity results in a hybridization of the cavity and waveguide modes into (almost) evenly spaced modes separated by the free spectral range (FSR) of the Fabry-P\'{e}rot cavity. The FSR is determined by the length of the waveguide and by the group velocity of the phonons. We design the waveguide to be single mode for the symmetry of the mechanical mode used in this work (the band structure is shown in Fig.~\ref{Fig:1}b).

\begin{figure}[t]
	\includegraphics[width = 1.\linewidth,right]{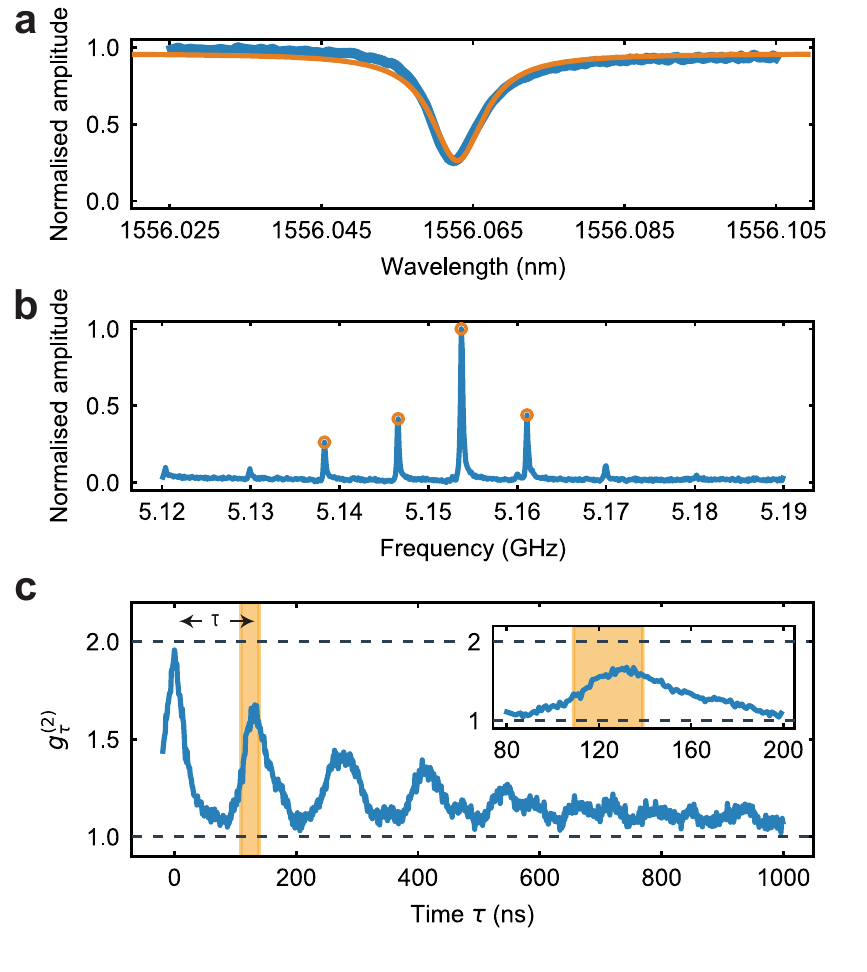}
	\caption{\textbf{Initial characterization.} a) Optical spectrum of the device measured in reflection. b) Mechanical spectrum measured using optomechanically induced transparency (OMIT). c) Second order correlations of a thermal state for different time delays between the SNSPDs click events $\tau$. The series of equally spaced peaks shows that when the thermal mechanical excitations leave the optomechanical cavity, they are reflected from the end of the waveguide and then return back into the cavity. The inset is a close-up of the area around the first peak. The shaded regions show the delay and control pulse area chosen for all the subsequent experiments in this work ($\tau = \SI{126}{n s}$, with a time length of \SI{30}{ns}).}
	\label{Fig:2}
\end{figure}

In order to create a propagating mechanical excitation we use a blue-detuned write (Stokes) control pulse, which via a two-mode squeezing interaction, creates entangled photon-phonon pairs. The phononic excitation created in the cavity then leaks into the waveguide, is reflected by the end mirror and returns back periodically to the cavity after a round-trip time $\tau$. Finally, to retrieve the mechanical state, a red-detuned read (anti-Stokes) control pulse enables the optomechanical beam-splitter interaction which maps the mechanical into a photonic excitation (see Fig.~\ref{Fig:1}c for details on the scheme). To create a time-bin encoded mechanical qubit using this scheme, we first place the device in a dilution refrigerator at \SI{10}{mK}, initializing the mechanical mode in its quantum ground state. We then send two blue-detuned write pulses separated by $\tau/2$ to the device in the cryostat, as shown in Fig.~\ref{Fig:1}d, and send the resulting scattered photons into an optical interferometer. One arm is delayed with respect to the other one by $\tau/2$ to overlap the scattered photons in time. The reflected control pulses are suppressed using optical filters (see SI) and the resulting interferometer output signals are detected on two superconducting nanowire single-photon detectors (SNSPDs). By operating in the low pulse energy regime (low optomechanical scattering probabilities, $p_{\text{w}}$), the two identical write pulses create the optomechanical state
\begin{align*}
	\ket{\psi_0} \propto & \ket{0000} + \sqrt{p_{\text{w}}} \big( \ket{1010}_{E_o L_o E_m L_m}+\\
	&\qquad e^{i\phase{w}} \ket{0101}_{E_o L_o E_m L_m} \big) + \mathcal{O}(p_{\text{w}}),
	\numberthis\label{Eq:1}
\end{align*}
where $E_m$ ($L_m$) and $E_o$ ($L_o$) indicate the ``Early" (``Late") mechanical and optical state, and $\phase{w}$ is the phase difference between the two ``Early" and ``Late" write pulses, set with an electro-optical modulator (EOM) (see SI section 4 for more details). By overlapping these ``Early" and ``Late" photons on a beamsplitter, after passing through the unbalanced interferometer, we erase any ``which path" information. Consequently, by detecting a Stokes-scattered photon from the overlapped ``Early" and ``Late" pulses on one of the detectors, we perform an entanglement swapping operation resulting in an heralded entangled state between the ``Early" and ``Late" traveling mechanical excitations
\begin{equation}
	\ket{\psi_m} \propto \ket{10}_{E_m L_m} \pm ~e^{i(\phase{w} + \phase{\text{off}})} \ket{01}_{E_m L_m},	
	\label{Eq:2}
\end{equation}
with the plus (minus) sign resulting from a detection event in either detector. The phase $\phase{\text{off}}$ is a fixed phase difference between the two arms of the unbalanced interferometer (SI sections 5 and 6). Note how the state is maximally entangled in the Fock-basis, which at the same time serves as a mechanical qubit.

The entangled phonon state travels through the waveguide and after re-entering the cavity can be mapped onto an optical state with the red-detuned anti-Stokes control pulses. The entire 4-mode optical state of write and read scattered photons can be expressed as:
\begin{align*}
	\ket{\psi_{AB}} \propto & \big[ (1+e^{i(\phase{w} + \phase{r} + 2\phase{\text{off}})})(\hat{a}_{\text{w},1}^{\dagger} \hat{a}_{\text{r},1}^{\dagger}  -  \hat{a}_{\text{w},2}^{\dagger}  \hat{a}_{\text{r},2}^{\dagger}) + \\
	&i(1-e^{i(\phase{w} + \phase{r} + 2\phase{\text{off}})})( \hat{a}_{\text{w},2}^{\dagger} \hat{a}_{\text{r},1}^{\dagger}  + \hat{a}_{\text{w},1}^{\dagger}   \hat{a}_{\text{r},2}^{\dagger}) \big]\ket{0000}
	\numberthis
	\label{Eq:3}
\end{align*}
where $\hat{a}_{\text{w},1}^{\dagger}$ ($\hat{a}_{\text{w},2}^{\dagger}$) and $\hat{a}_{\text{r},1}^{\dagger}$ ($\hat{a}_{\text{r},2}^{\dagger}$) are the creation operators of the photon coming from the write pulse and read pulse, on detector 1 (2). An additional phase $\phase{r}$ is applied only on the ``Late" read pulse, which is used to rotate the readout basis. Here the read pulses map the mechanical state onto the optical mode and hence this state is a direct result of the entanglement between the photonic and the traveling phononic qubit. For verifying the mechanical entanglement of Eq.~\ref{Eq:2} we use $\phase{r} = 0$.

In order for our protocol to work, we need to fulfill several basic requirements. For both the phononic and photonic qubits, for example, we have to create orthogonal states (the basis), and thus for the time-bin encoding we have to be able to unambiguously distinguish the ``Early" and ``Late" states. 
Experimentally we implement this by realizing a $\sim$\SI{100}{\micro m} long phononic waveguide and by choosing the control pulse length as \SI{30}{ns}, given a simulated group velocity in the waveguide of approx.\ $\SI{2000}{m/s}$. Moreover the thermal occupation of the mechanical mode, mainly given by a small absorption of the control pulses in the optomechanical cavity, has to be $\ll 1$ in order to realize a high-fidelity entangled state. This limits the maximum scattering probability of the write and read control pulses (see SI section 3).

To characterize the device we first measure its optical properties at \SI{10}{mK} and observe a resonance at $\lambda \approx \SI{1556.06}{nm}$ with FWHM of $\kappa/2\pi \approx \SI{1.05}{GHz}$ (intrinsic loss rate $\kappa_\text{i} /2\pi= \SI{250}{MHz}$, see Fig.~\ref{Fig:2}a). We use the optomechanically induced transparency (OMIT) technique to measure the mechanical spectrum of the device~\cite{Weis2010,Zivari2022}. As can be seen in Fig.~\ref{Fig:2}b, the hybridized modes exhibit a clean, evenly spaced spectrum with $\textrm{FSR}\approx\SI{8}{MHz}$. We choose the most prominent mechanical resonance in Fig.~\ref{Fig:2}b (around \SI{5.154}{GHz}) as the frequency to which we detune the lasers with respect to the optical resonance in order to address Stokes and anti-Stokes interactions. We further use the rate of Stokes-scattered photons from a \SI{30}{ns} long pulse to determine the equivalent single photon optomechanical coupling rate~\cite{Stockill2019} of the ensemble of optomechanically coupled modes at $g_{0}/2\pi\approx$~\SI{380}{kHz}.

In order to measure the round-trip time and coupling between the cavity and waveguide we pump our device with a continuous red-detuned laser. Due to the non-zero optical absorption in the device, the continuous laser creates a thermal mechanical population inside the optomechanical device that leaks into the phononic waveguide and reflects from the free standing end before returning back into the optomechanical cavity. The same red-detuned laser then maps the mechanical state onto a photonic state and we measure the two-photon detection coincidence with varying delays between two events. This measurement allows us to obtain the intensity correlation $g^{(2)}_{\tau}$ of the mechanical thermal state in the optomechanical cavity, as can be seen in Fig.~\ref{Fig:2}c. 
The coincidence rate is normalized to the single photon click rates. For zero time delay we find a $g^{(2)}_{\tau=0} \approx 2$, as expected for a thermal state. The correlation drops down to 1, as the thermal state leaves the optomechanical cavity into the waveguide resulting in uncorrelated clicks, and then periodically increases again when the thermal population returns back to the cavity~\cite{Zivari2022}. We use this measurement to determine the round-trip time for an excitation in the waveguide, $\tau = $~\SI{126}{ns}.

\begin{figure}[H]
	\includegraphics[width = .9\linewidth,right]{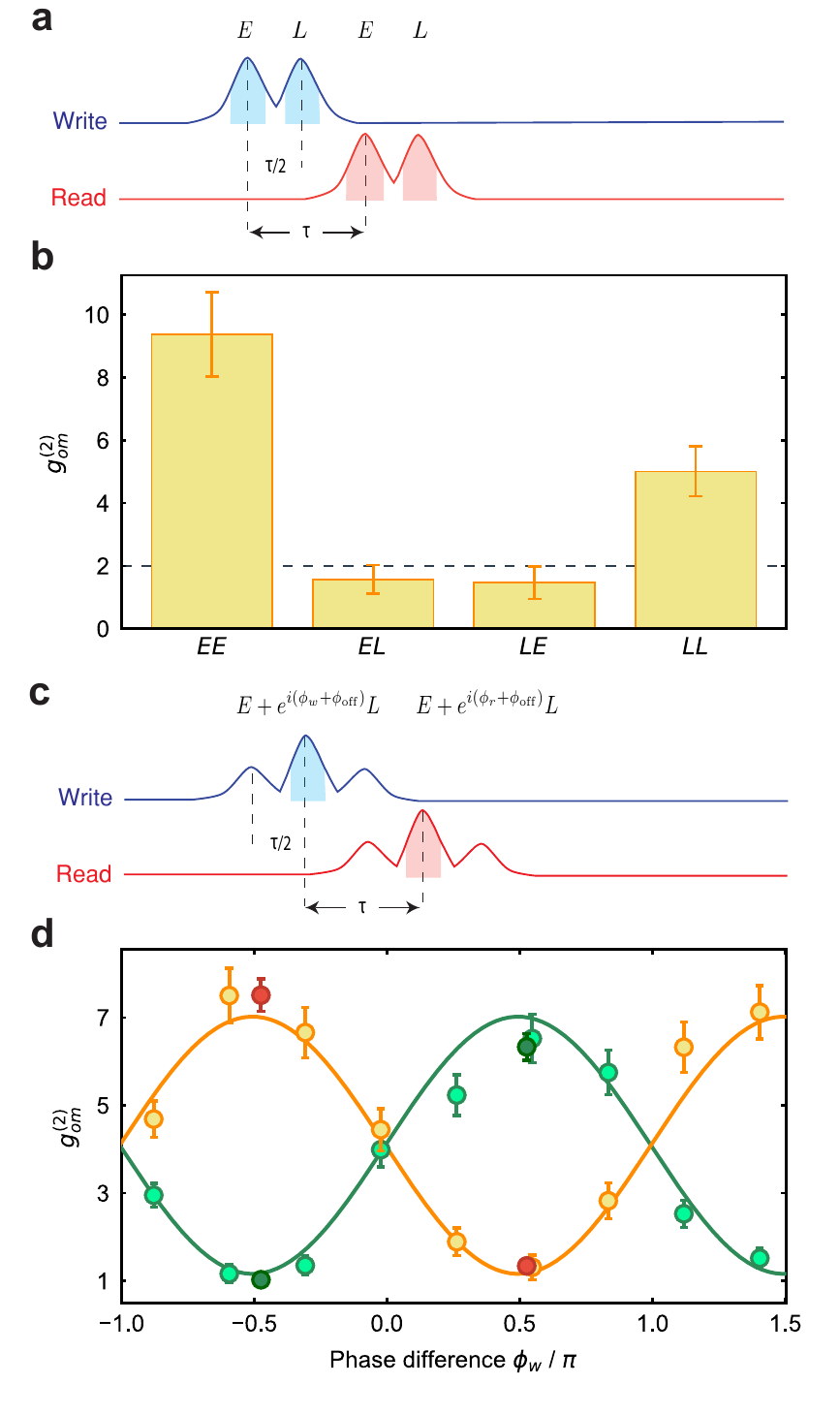}
	\caption{\textbf{Time-bin phononic entanglement.} a) Pulse scheme for a double write / read pulse cross-correlation measurement. In this experiment the delayed arm of the interferometer is open. The two write and two read pulses are called ``Early" ($E$) and ``Late" ($L$) and are delayed by $\tau/2$. The shaded areas are the regions from which the coincidences are gathered (time length of \SI{30}{ns}). b) Extracted values of the cross-correlation $g^{(2)}_\text{om}$ measured for the four combinations of write / read pulses. The correlations for $EE$ and $LL$ are significantly exceeding the classical threshold of 2 (dashed line), while the other two combinations of $EL$ and $LE$ only exhibit classical correlations. c) Control pulse scheme to create and detect time-bin phononic entanglement at the interferometer. The ``Early" pulse passing through the delay arm of the interferometer and the ``late" pulse passing through the direct arm of the interferometer overlap in time. d) Second-order correlations of the Stokes and anti-Stokes photons as a function of the relative phase difference $\phase{w}$ between the ``Early" and ``Late" write pulses. The events for same detector coincidences are shown in green, while different detectors coincidences are orange. Two additional measurements, red and dark green points, are performed at two phase settings to obtain more statistics for verifying the phononic entanglement. The maximum violation is $R=0.72\pm0.06$, almost 5 SDs below the classical threshold of 1. All error bars are one SD. The solid curves are the joint fit of the data and serve as guide to the eye.}
	\label{Fig:3}
\end{figure}

Note that the decay in the peak values is mainly due to the small difference in FSR between the mechanical modes~\cite{Zivari2022}, as well as the short mechanical lifetime $T_1 \approx$ \SI{2.2}{\micro s} (see section 7 in the SI for more details). From the FWHM of the peak centered around zero time delay in Fig.~\ref{Fig:2}c we can extract the packet time duration of $\approx\SI{30}{ns}$. To match the packet time length, we then choose to use \SI{30}{ns}-long Gaussian write and read pulses in all experiments.

As a next step we verify our ability to distinguish between multiple phonon wavepackets. To do this, we measure the photon cross-correlations in a double write / read pulse sequence, in which we create and measure the second wavepacket after half of the round trip time $\tau/2 \approx$~\SI{63}{ns}. In this experiment, the delay arm of the interferometer is disconnected, such that the pulses do not interfere. Two write and two read pulses are sent to the device with a relative delay between them of $\tau/2$, as shown in Fig.~\ref{Fig:3}a. In all the pulsed measurement we set a waiting time between trials of $\sim$7 $T_1$ (\SI{15}{\micro s}) to let the mechanical modes thermalize to the ground state (see SI section 3). The energy of each pulse is \SI{26}{fJ} (\SI{112}{fJ}) for the write (read), probabilistically scattering photons through the Stokes (anti-Stokes) process, with a probability of $p_{\text{w}} = 0.2\%$ ($p_{\text{r}} = 0.7\%$). The measured thermal phonon number of the mechanical resonator after applying the four pulses are $0.022\pm0.002$, $0.040\pm0.003$, $0.066\pm0.003$ and $0.095\pm0.004$ (cf.\ SI section 3). We measure the second order cross-correlation between the four combinations of ``Early" and ``Late" write and read pulses, $g^{(2)}_\text{cc}$ as shown in Fig.~\ref{Fig:3}b. We observe strong non-classical correlations of $g^{(2)}_\text{cc} = 9.4 \pm 1.3$ between ``Early"-``Early" and $g^{(2)}_\text{cc} = 5.0 \pm 0.8$ between ``Late"-``Late" combinations, while the other two combinations show only classical correlations of $g^{(2)}_\text{cc} = 1.5 \pm 0.5$~\cite{Hong2017}. Note that the lower value for the ``Late"-``Late" combination, with respect to ``Early"-``Early", is caused by the small accumulated thermal population induced by the pulses (see SI section 3 for more information).

We now proceed to verify that we have created a traveling mechanical qubit encoded in a superposition of ``Early" and ``Late" time bins, by sending the same pulse sequence to the device, with the delay arm of the interferometer connected. This way the part of the ``Early" scattered photons that pass through the delay line and the part of the ``Late" scattered photons that pass though the direct arm are overlapped in time, such that a single photon detection event after BS2 projects the mechanical state in Eq.~\ref{Eq:2}. The pulse sequence at the detectors is shown in Fig.~\ref{Fig:3}c, where the highlighted peaks are the overlapped and interfered ``Early" and ``Late" pulses, from which we detect the photons. We sweep the excitation phase $\phase{w}$ and measure the second order correlation $g^{(2)}$ between the write and read photon detection events occurring at the same (green) or different (orange) output of BS2, displayed in Fig.~\ref{Fig:3}d. The periodic dependence on the phase demonstrates the coherence of the generated entangled state. To show that the state shown in Eq.~\ref{Eq:2} is indeed entangled, we use an entanglement witness, denoted by $R$, designed for optomechanical systems~\cite{Borkje2011}, as previously used in~\cite{Riedinger2018}. After sufficient integration we gather more than 500 coincidence events and we obtain $R=0.72\pm0.06$, violating the classical threshold of 1 by almost 5 standard deviations.

\begin{figure}[]
	\includegraphics[width = 1\linewidth]{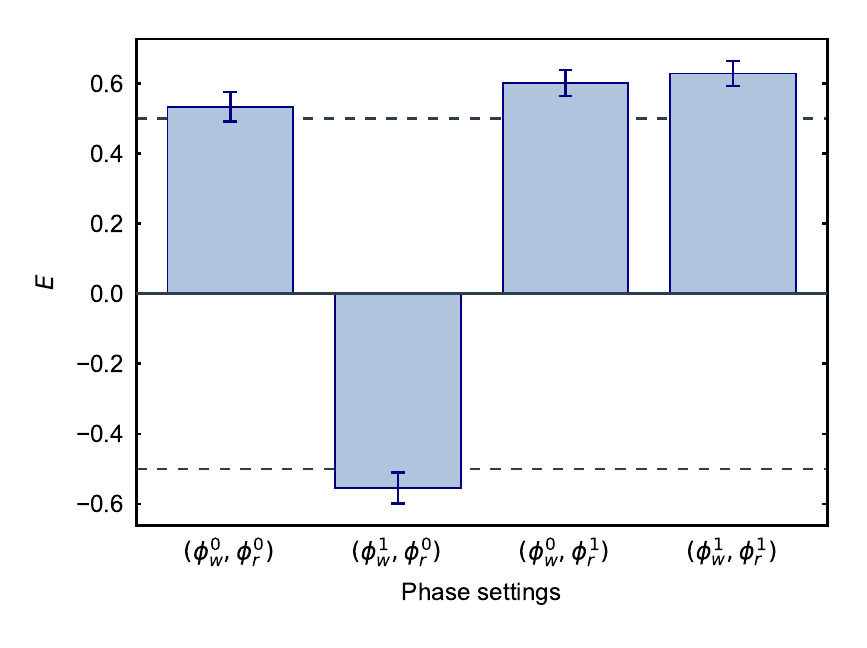}
	\caption{\textbf{Bell test.} The values of the correlation coefficients $E$ grouped for the four ideal phase settings of the CHSH inequality, which are $(\pphase{w}{0}, \pphase{r}{0}) = (\phase{0}-\pi/4$, $0$), $(\pphase{w}{1}, \pphase{r}{0}) = (\phase{0}+\pi/4$, $0$), $(\pphase{w}{0}, \pphase{r}{1}) = (\phase{0}-\pi/4$, $\pi/2$) and $(\pphase{w}{1}, \pphase{r}{1}) = (\phase{0}+\pi/4$, $\pi/2$). The total number of events for each phase setting is $\sim$400. The dashed lines are the threshold for each correlation coefficient to violate the inequality. From these values we obtain $S=2.32\pm0.08$, which violates the inequality by 4 SD. All errors are one SD.}
	\label{Fig:4}
\end{figure}

To unambiguously demonstrate the non-classical character of the traveling phononic qubit and the photon state in Eq.~\ref{Eq:3}, we perform a Bell-type test using the CHSH inequality~\cite{Clauser1969}. We define the correlation coefficients
\begin{equation}
	E(\phase{w},\phase{r}) = \frac{n_{11}+n_{22}-n_{12}-n_{21}}{n_{11}+n_{22}+n_{12}+n_{21}},
\end{equation}
where $n_{kl}$ are the events where detector $k$ clicked after a write pulse (station A in Fig.~\ref{Fig:1}d) and detector $l$ after a read pulse (station B in Fig.~\ref{Fig:1}d). The inequality then states that
\begin{equation}
	S = |E(\pphase{w}{0},\pphase{r}{0})-E(\pphase{w}{1},\pphase{r}{0})+E(\pphase{w}{0},\pphase{r}{1})+E(\pphase{w}{1},\pphase{r}{1})| \leq 2.
\end{equation}
The maximum violation is expected to occur for ($\pphase{w}{i}=\phase{0}+(-1)^{i+j}\pi/4$, $\pphase{r}{j}=(\pi/2)^j$), with $i,j = \{0,1\}$, and where $\phase{0}=2\phase{\text{off}} + \pi/2 \approx 1.0\pi$ is the phase for which the correlation coefficient is zero (with negative slope). We choose phase settings with a small offset compared to these values to have the highest possible value of $S$ for our setup (see SI section 6 for more details). To violate the CHSH inequality we lower the energy of the write pulses slightly, such that we use \SI{15}{fJ} (\SI{112}{fJ}) for the write (read) pulse, with a scattering probability of  $p_{\text{w}} = 0.13\%$, ($p_{\text{r}} = 0.7\%$). The measured thermal populations for the four pulses are then $0.027\pm0.003$, $0.038\pm0.004$, $0.055\pm0.002$ and $0.090\pm0.004$ (see SI section 3).

From the correlation coefficient we define the visibility as $V=max(|E|)$. We first perform an additional measurement at the phase of maximum visibility obtaining $V=0.82\pm0.04$, which is considerably higher than the threshold of $V>1/\sqrt{2}\approx0.7$ required for violating the CHSH inequality. We then measure at the four optimal phase settings for the Bell test (see Fig.~\ref{Fig:4}) obtaining a value of $S=2.32\pm0.08$, which corresponds to a violation of the CHSH inequality by 4 standard deviations. The rate of events for this measurements is around 30 per hour of integration, allowing us to measure the full data set for the violation within 56 hours.

\section{Discussion}

We have unambiguously demonstrated a traveling phononic qubit in the form of a time-bin entangled state, which can be used to distribute quantum information on a chip. The routing process is shown to be fully coherent, which is of fundamental importance for routing quantum information and interconnecting quantum devices. While we limit ourselves to show two-mode entanglement, the same device can be used with up to four modes, given the round-trip time and mechanical packet length, or more by using a longer waveguide. Additionally, the quantum state can be retrieved at arbitrary multiples of the round trip time, allowing for long storage and controlled emission of the state. Moreover, as the phononic entangled state travels down the waveguide, a straightforward extension using our device will allow to distribute quantum entanglement to different points on a chip. We have chosen to use a waveguide design with a lifetime of only $T_1 \approx$ \SI{2.2}{\micro s}, limiting the maximal phonon traveling length for this devices to around \SI{3}{mm}. By adding additional phononic shielding, this can however easily be extended to meter scales as the device's lifetime increases to several milliseconds~\cite{Wallucks2020}.

The demonstrated time-bin entanglement between a photonic and a traveling phononic qubit, verifying their non-classical correlations by violating a CHSH inequality, underlines the suitability of the phononic system as a DLCZ unit cell~\cite{Duan2001}. In this work, the fidelity of the entangled state is limited by residual optical absorption, which can be further reduced by up to an order of magnitude through optimized fabrication, allowing for state retrieval efficiencies of up to \SI{30}{\%}~\cite{Hong2017}.

The ability to excite, guide, and detect traveling phonons is the basic toolbox for phonon manipulation on-chip, enabling a completely new field using traveling mechanical modes in the quantum regime. Together with a phononic phase modulator~\cite{Taylor2022} and beamsplitter, this work will lead to full coherent control of guided phonons and paves the way to novel quantum acoustic experiments. Moreover, our measurements highlight the potential of phonons as ideal candidates for realizing quantum networks and repeaters, as well as for on-chip distribution of quantum information in hybrid quantum devices, for example for interfacing microwave superconducting circuits with spin quantum memories~\cite{Neuman2021} or to couple on-chip qubits using electron-phonon interaction in solids~\cite{McNeil2011}.

\section{Methods}

The device is fabricated from a silicon-on-insulator (SOI) wafer with a \SI{250}{nm} thick silicon device layer. We use electron beam lithography to pattern the structure and transfer the mask with a dry HBr/Ar plasma etch. The chip is diced in order to access the device's optical waveguide with a lensed fiber in the dilution refrigerator. Finally, the device is cleaned with a piranha solution and released using a 40\% hydrofluoric acid (HF) wet etch to remove the oxide layer~\cite{Riedinger2016}. After room temperature characterization and just before the cooldown in the cryostat, we perform another piranha cleaning and a 1\% HF dip in order to minimize the native oxide.

\section*{Acknowledgments}
\begin{acknowledgments}
	We would like to thank Matteo Lodde, Andrea Fiore and Bas Hensen for valuable discussions. We further acknowledge assistance from the Kavli Nanolab Delft. This work is financially supported by the European Research Council (ERC CoG Q-ECHOS, 101001005) and is part of the research program of the Netherlands Organization for Scientific Research (NWO), supported by the NWO Frontiers of Nanoscience program, as well as through Vidi (680-47-541/994) and Vrij Programma (680-92-18-04) grants. R.S.\ acknowledges funding from the European Union under a Marie Sk\l{}odowska-Curie COFUND fellowship.
\end{acknowledgments}

\textbf{Conflict of interests:}\ The authors declare no competing interests.

\textbf{Author contributions:}\ A.Z., R.S. and S.G.\ devised and planned the experiment. A.Z., N.F, R.B.\ simulated and designed the device. N.F.\ fabricated the sample, A.Z. and N.F.\ built the setup and performed the measurements. A.Z., N.F., R.S. and S.G.\ analyzed the data and wrote the manuscript with input from all authors. E.V. and S.G. supervised the project.

\textbf{Data Availability:}\ Source data for the plots are available on \href{https://doi.org/10.5281/zenodo.7123020}{Zenodo}.

\setcounter{figure}{0}
\renewcommand{\thefigure}{S\arabic{figure}}
\setcounter{equation}{0}
\renewcommand{\theequation}{S\arabic{equation}}

\clearpage

\section*{Supplementary Information}
\label{SI}

\subsubsection{Design}
\label{SI:design}
\begin{figure*}
	\includegraphics[width = 0.9\linewidth]{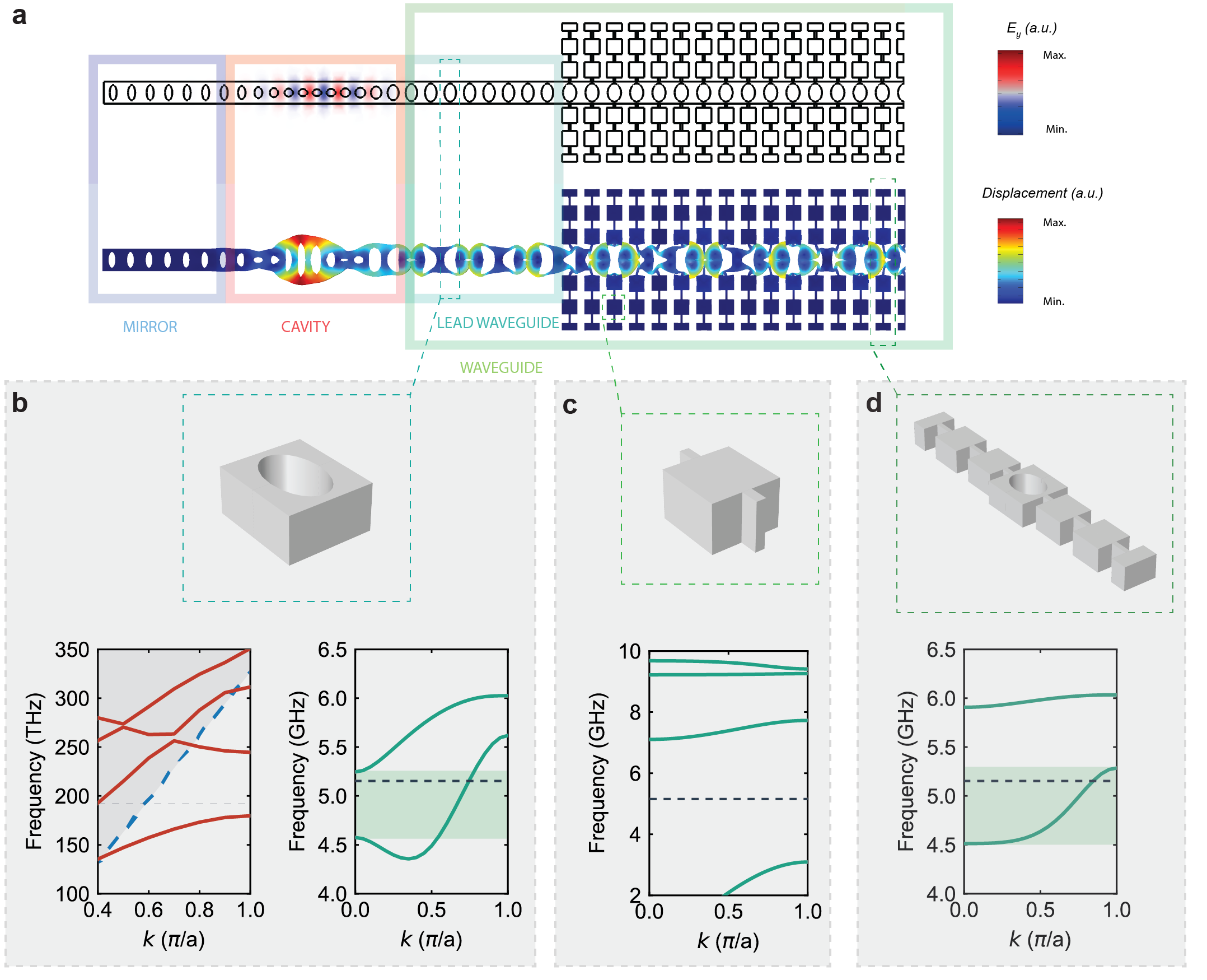}
	\caption{a) Optical (top) and mechanical (bottom) eigenmode simulations of the full structure. The mechanical mode shown is a cavity-waveguide supermode. Note how the optical mode is confined in the cavity, while the mechanical mode is extended into the waveguide. b) Lead waveguide unit cell, with its optical (left) and mechanical (right) band diagram shown below. The gray area for the optical part depicts the light cone, delimited by the blue dashed line. c) Shield unit cell, used in the clamps to connect the waveguide to the substrate and its mechanical band diagram below. d) Unit cell of the shield-clamped waveguide, with its band structure shown underneath. In all plots the horizontal black dashed lines are the working optical and mechanical frequencies, while the highlighted area are the single mode regions of the waveguides.}
	\label{Fig:S1}
\end{figure*}

Our device, which is shown in~Fig.~\ref{Fig:1}a, is composed of three distinct parts:\ a mirror, a cavity and a waveguide. To design the device we use finite element simulations (COMSOL) and engineer a suspended silicon nanobeam (width \SI{529}{nm} and thickness of \SI{250}{nm}) with elliptical holes patterned into it. The hole dimensions are varied along the beam in order to realize the different parts of our device. The finite element simulation of the full structure is shown in Fig.~\ref{Fig:S1}a. The left part (blue) is a phononic and photonic mirror, with the respective bandgaps at the resonance frequencies of the cavity. The optomechanical cavity, acting as the source and detector for phonons, has a co-localized single optical mode in the telecom band and a single mechanical mode around \SI{5}{GHz}, similar to~\cite{Chan2011,Riedinger2016}. The phononic waveguide (green) is a single mode phononic waveguide for the symmetric mechanical breathing mode in the frequency range of interest (around $\SI{5}{GHz}$). Only this mode is considered, as it matches the mechanical mode shape of the cavity, enabling large mechanical coupling between the cavity and the waveguide. The first part of the waveguide, referred to as the ``lead waveguide" in the figure, acts as a photonic mirror having a bandgap in the telecom range. The unit cell of this part, together with its optical and mechanical band structure, are shown in Fig.~\ref{Fig:S1}b and c, respectively.  As a result of our design, the optical mode stays confined inside the optomechanical cavity, while the mechanical mode is guided with very little loss through the waveguide, as shown in Fig.~\ref{Fig:S1}a. The second part of the waveguide is connected to the substrate via phononic shield clamps for structural support. The band diagram of the phononic shield together with the unit cell are shown in Fig.~\ref{Fig:S1}d and e. The shield minimizes the mechanical loss from the waveguide to the substrate, fully confining the mechanical mode in the waveguide. The design of the waveguide with clamping differs from the lead waveguide to avoid perturbations of its band structure. The unit cell shape of this waveguide and the band structure are shown in Fig.~\ref{Fig:S1}f and g. Note how the mode in the waveguide has an approximately linear dispersion in the range of interest.

\subsubsection{Mechanical Lifetime}
\label{SI:lifetime}
\begin{figure}[h]
	\includegraphics[width = 1\linewidth]{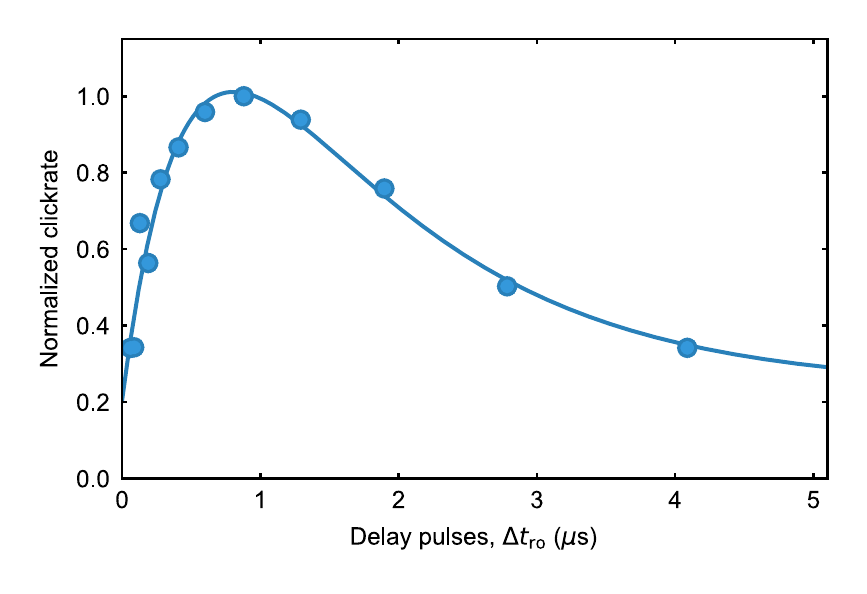}
	\caption{Normalized clickrates from the probe pulse that gives an uncalibrated measure of the thermal population in time. The delay between pump and probe pulse is $\Delta t_\text{ro}$. We extract $T_1 \approx \SI{2.2}{\micro s}$, by fitting the data (solid curve).}
	\label{Fig:S2}
\end{figure}

To measure the mechanical lifetime of the device ($T_1$) we send a series of red-detuned double pulses with the interferometer delay arm open. The strong first pulse creates, via optical absorption, a relatively large thermal population that is probed by the second pulse, and which is delayed by time $\Delta t _{ro}$. With this pump-probe experiment we can access the uncalibrated thermal population of the device in time~\cite{Hong2017} (see Fig.~\ref{Fig:S2}). From the exponential decay we measure $T_1 \approx \SI{2.2}{\micro s}$, much longer than the $\SI{126}{n s}$ delay used in the experiments. We set the time between trials in all experiments equal to $\SI{15}{\micro s}$, to let the population fully decay. Note that while the device is intentionally designed to have a short lifetime in order to allow for a high repetition rate of the experiment, previous work with similar structures has reported lifetime up to 5.5~ms~\cite{Zivari2022}. We would also like to note that an additional phononic shield period at the mirror side (blue part in Fig.~\ref{Fig:1}a) does not increase the lifetime any further. The increase in thermal population for short delays ($\Delta t _{ro}<\SI{1}{\micro s}$) is given by the delayed absorption~\cite{Riedinger2016}.

\subsubsection{Thermal occupancy of the mechanical mode}
\label{SI:thermal_occupancy}
\begin{figure*}
	\includegraphics[width = 1\linewidth]{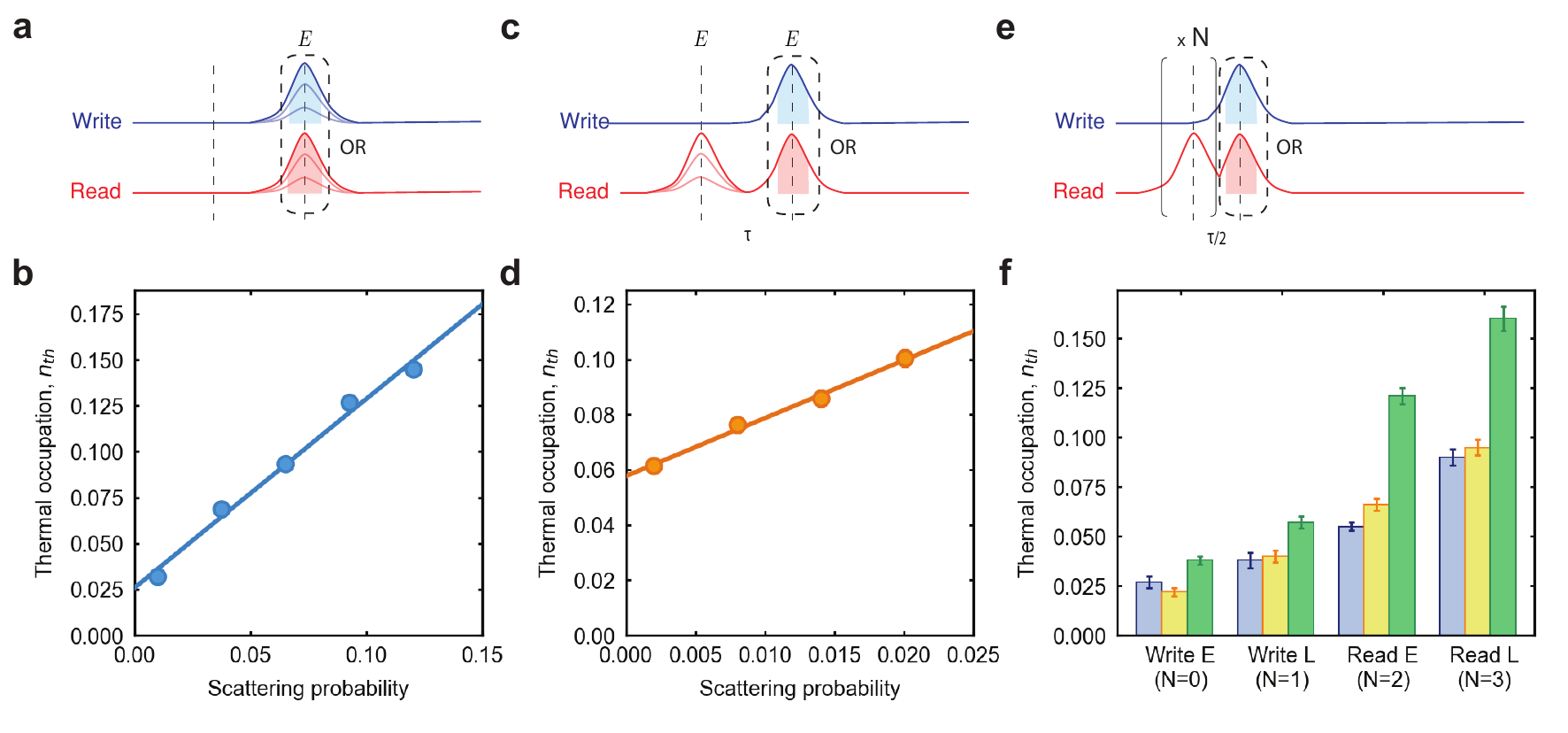}
	\caption{a) Pulse scheme for the thermal occupation measurement. We send trains of pulses alternating between write and read to measure the asymmetry in their scattering probabilities. The opaque pulses represents the sweep in energy of the pulse, while the shaded area is the integration region. b) Thermal occupation as a function of the measured scattering probability for a single pulse (equivalent to the read ``Early" situation in the experiments). The solid line is a linear fit to the data. c) Same as a) for the two pulse calibration of the thermal occupation. Note how the read pre-pulse is only for heating the mechanical mode and its energy is swept. The alternating pulses used to measure the $n_\text{th}$ are in the black dashed box and have fixed energy. d) Thermal occupation of the second pulse (read ``Early" in the experiment) as a function of the measured scattering probability of the first pulse (write ``Early" in the experiment). Note how the offset in this measurement strongly depends on the energy of the second pulse, which in this case is \SI{225}{fJ} (twice the energy of the single read pulses used in the phononic entanglement experiment of Fig.~\ref{Fig:3} in the main text). The solid line is a linear fit to the data. e) Same as a) but for multiple pre-pulses. We send $N$ pre-pulses (with $N=(0,1,2,3)$ for write E, write L, read E and read L, respectively) and use the alternating pulses in the dashed black box to measure the thermal population. Each pulse has the energy used in the experiments. f) Measured thermal occupation for the four pulses used in the experiments. In blue for the scattering probabilities used in the measurements for the additional phononic entanglement data and Bell test (Fig.~\ref{Fig:4} in the main text), in orange for the double pulse cross correlation and the phononic entanglement with the sweep in $\phase{w}$ (Fig.~\ref{Fig:3} in the main text) and in green for the phase calibration (section~\ref{SI:phase_calibration}). All error bars are one standard deviation.}
	\label{Fig:S3}
\end{figure*}

In order to determine the thermal occupation of the mode of interest we send trains of alternating write and read pulses to the device, as shown in Fig.~\ref{Fig:S3}a. From the asymmetry in clickrates of these pulses we can calculate the thermal mechanical population $n_\text{th}$. We adjust the scattering probability by sweeping the energy of the pulses and measure the resulting thermal occupation, see Fig.~\ref{Fig:S3}b. To further mimic the same heating conditions as in the experiment, without the optomechanically excited coherent population created from the write pulses, we use heating pre-pulses from the read laser. The alternating pulses, used to measure $n_\text{th}$, are sent at a delay given by the round-trip time $\tau$ and have a fixed energy, while the energy of the pre-pulse is swept. In Fig.~\ref{Fig:S3} we show the pulse scheme for these measurements, as well as the thermal occupation at the second pulse (in the experiment the read pulse) as a function of the scattering probability of the first pulse (in the experiment the write pulse). We use the two asymmetry measurements (Fig.~\ref{Fig:S3}b and d) to choose the single write/read scattering probabilities that will give a total thermal population below $0.1$, with a third of this thermal occupation given by the first pulse and the rest by the second (to minimize the effects of delayed heating).

We then measure the values of $n_\text{th}$ for the pulses used in the experiments. We send heating pre-pulses and use the alternating pulses to measure the thermal population, as drawn in Fig.~\ref{Fig:S3}e. Each of these pulses has the corresponding energy used in the experiments and all are delayed by $\tau/2$ with respect to each other. In Fig.~\ref{Fig:S3}f we report the measured thermal occupation for the four pulses of the experiments, for the three sets of scattering probabilities used:\ in blue the one for the additional phononic entanglement data and the Bell test (Fig.~\ref{Fig:4} in the main text), in orange for the double pulse cross correlation and the phononic entanglement with the sweep in $\phase{w}$ (Fig.~\ref{Fig:3} in the main text) and in green for the phase calibration (section~\ref{SI:phase_calibration}). Note how the thermal population increases non-linearly with increasing number of pulses due to delayed heating, as is clearly visible in Fig.~\ref{Fig:S3}.

\subsubsection{Experimental setup}
\label{SI:setup}
\begin{figure*}
    \includegraphics[width = 1\linewidth]{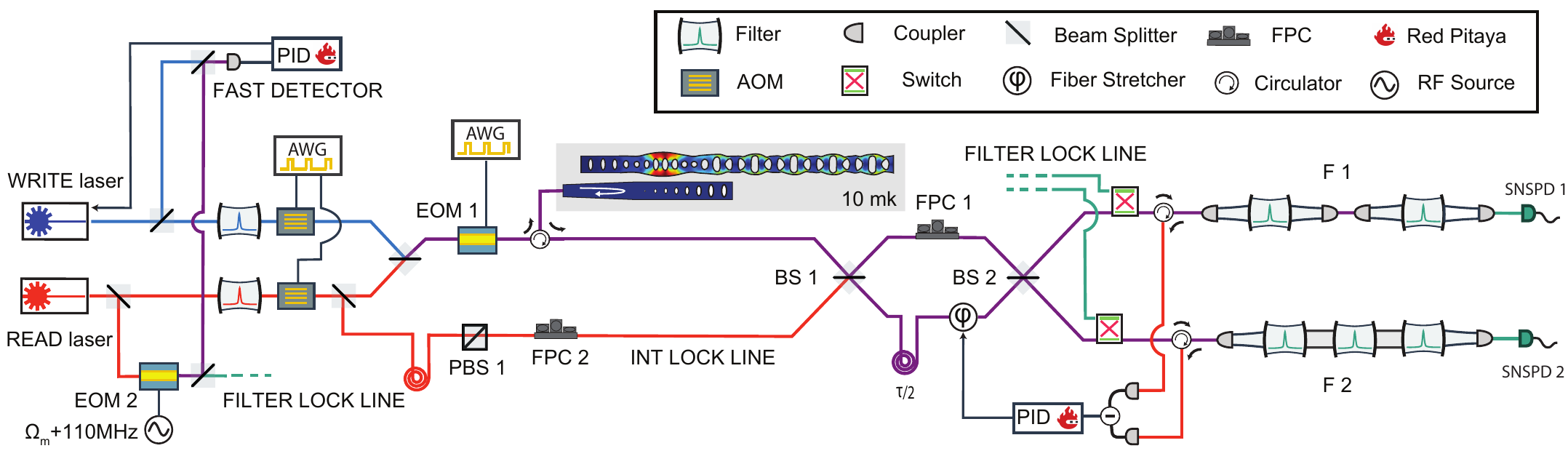}
    \caption{Detailed scheme of the setup (see text for more details). AOM are the acousto-optic modulators, AWG
    	the arbitrary waveform generators, EOM the electro-optic modulator, VOA the variable optical attenuator, BS the
    	beamsplitters, and PBS the polarizing BS, FPC the fiber polarization controller and SNSPD the superconducting nanowire single-photon detectors. $\Omega_\text{m}$ the mechanical frequency. All the components (apart from the free-space filters F 1 and F 2) are fiber based.}
    \label{Fig:S4}
\end{figure*}

A sketch of the experimental setup is shown in Fig.~\ref{Fig:S4}. Two continuous-wave (CW) lasers (write and read) are frequency-locked to one another by detecting the interference between their light on a fast detector (in particular:\ the light from the write laser and the second order sideband of the read laser generated by EOM 2, which have a frequency difference of 2$\times$\SI{110}{MHz}). The laser light from the CW lasers is filtered to remove GHz noise using fiber filters. The pulses are created by gating the CW light with \SI{110}{MHz} AOMs, which are driven by an arbitrary waveform generator and the laser pulses are combined on a beam splitter. A phase EOM, driven by another AWG, is used to add a phase offset to the "Late" pulses (to set $\phase{r}$ and $\phase{w}$). The pulses are then routed to a circulator and to the cryostat, where a lensed fiber allows the coupling to the device's optical waveguide, with efficiency of $\eta_\text{c} \approx 50 \%$. The light from the device is fed to an unbalanced Mach-Zehnder interferometer, defined by BS 1 and BS 2, where the time delay between the arms is $\tau/2=\SI{63}{n s}$. These two BSs have a relative difference in the splitting ratio of the two output ports smaller than 0.5\%, while the losses in both lines are negligible. The interferometer is actively stabilized using a home-built fiber stretcher controlled by a PID loop that uses the signal from pulses coming from the INT LOCK LINE (see section~\ref{SI:phase_stability} for more details). The polarization of the two arms are matched at BS 3 using the fiber polarization controller 1 (FPC 1). The light from the interferometer is filtered by two sets of free space optical Fabry-P\'erot cavities (F 1 and F 2) with suppression ratios of the strong control pulses of about 115~dB (F 1) and 112~dB (F 2). This gives a pump pulse leakage rate of $2\times 10^{-7}$ and $4\times 10^{-7}$ photons per repetition from the write pulse and $1.4\times 10^{-6}$ and $2.6\times 10{-6}$ from the read pulse, for F 1 and F 2, respectively. Note that the two sets of filters have a CW efficiency of transmission at resonance of $\sim$65\%. Due to the different total bandwidth of \SI{40}{MHz} (\SI{80}{MHz}) for F 1 (F 2), the relative transmission efficiency of the pulses is about  40\% lower for F 1 using \SI{30}{ns} long pulses. The experiment is paused and the filters are locked on resonance with the cavity every \SI{8}{s}, flipping the switches to use the CW signal from the FILTER LOCK LINE (detectors not shown). The average time needed to lock the two filter setups is about $\SI{1}{s}$. The signal photons are detected using superconducting nanowire single photon detectors (SNSPD).

\subsubsection{Phase stability}
\label{SI:phase_stability}
A fundamental part of the experiment is the phase difference acquired by the pulses in the unbalanced Mach-Zehnder interferometer, which has to be actively locked. In order to do so, the two strong control pulses from the read laser line are routed via a 90:10 BS to a long delay line ($\approx$\SI{1}{\micro s} of delay) to have them temporally spaced from the signal. A PBS is inserted in the line to minimize polarization drifts. After passing through the unbalanced interferometer, the pulses are reflected by the first cavity of both filter setups and are routed by circulators to a balanced detector. A sample\&hold board (not shown in Fig~\ref{Fig:S4}) is used to select the correct pulse and feed it as the input voltage to a Red Pitaya board. The output of the board is amplified (not shown) and sent to the fiber stretcher. The Red Pitaya runs a PID program~\cite{Luda2019} and the feedback loop is ultimately limited by the bandwidth of the fiber stretcher (approx.\ \SI{20}{kHz}).

\begin{figure*}[ht]
	\includegraphics[width = 1\linewidth]{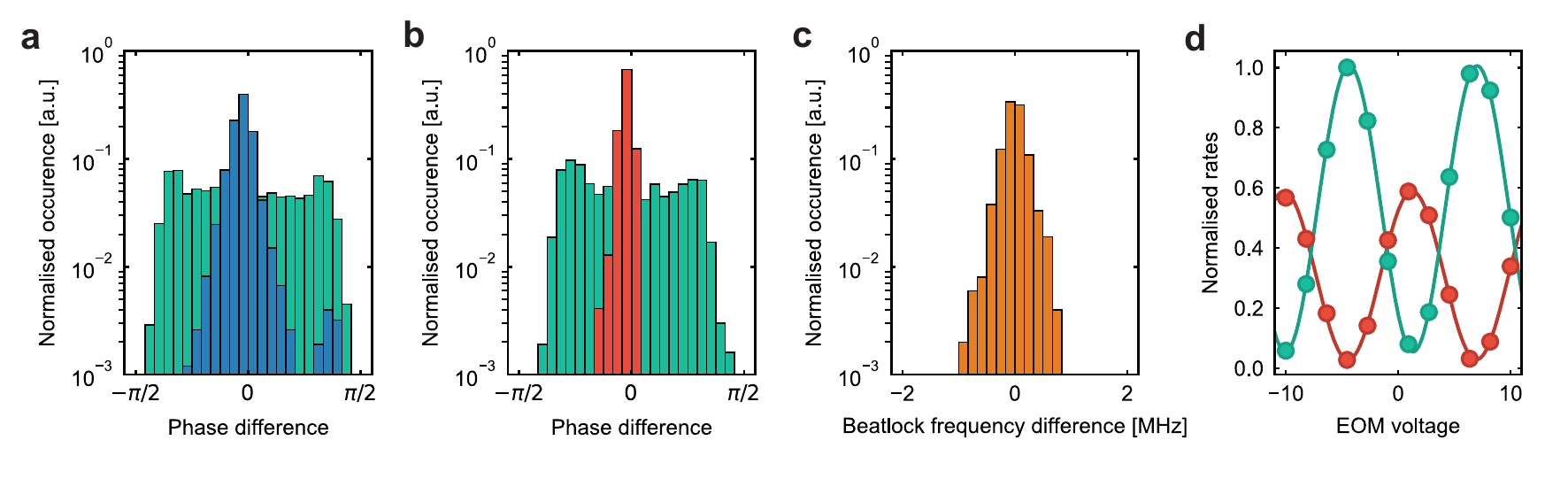}
	\caption{ a) Occurrence histogram of the phase difference acquired by the write pulses when the interferometer is locked (in blue) and unlocked (in green). The FWHM is $\approx \pi/7$. b) Same as a) for the read pulses, for the case of a locked interferometer (in red) and unlocked (in green). The FWHM here is $\approx \pi / 20$. c) Occurrence histogram of the frequency difference of the write and read lasers as from the beatlock. Here the FWHM is $\approx\SI{0.5}{MHz}$. This relative frequency jitter causes a bigger phase difference spread for the write pulses compared to the read ones. d) Normalized count rates of SNSPD1 (red) and SNSPD2 (green) for weak coherent pulses on resonance with the filter cavities. Sweeping the EOM voltage results in the observed interference pattern, which we use to calibrate the phase given by the EOM. In this case the interferometer visibility is $V_{\text{int}}\approx\SI{94}{\%}$. Error bars are one standard deviation and are too small to be seen.}
	\label{Fig:S5}
\end{figure*}

The phase stability can be measured by tracking the voltage of the locking pulse on the balanced detector. In Fig.~\ref{Fig:S5}a (b) we plot the occurrence histogram of the phase difference acquired passing the interferometer for the write (read) pulses, in the case the interferometer is locked (in blue (red)), or unlocked (in green). The FWHM are $\approx \pi/7 $ for the write and $\approx \pi/20 $ for the read pulses. This phase spread is the same for the Stokes (anti-Stokes) scattered photons. Note that the FWHM of phase difference for write and read pulses are different, since the phase acquired by the write laser pulse also depends on the relative frequency jitter of the two lasers. This will only affect the pulses from the write laser since the lock pulses are generated from the read laser. This frequency jitter is reported in Fig.~\ref{Fig:S5}c, where the occurrence histogram of the frequency difference from the beatlock is shown. Here the FWHM is around $\SI{0.5}{MHz}$. 

We use FPC 2 to balance the lock signal from the INT LOCK LINE at the balanced detector. We then lock the interferometer and use the first order interference from very weak pulses from the write laser, on resonance with the filter cavities, to measure the interferomenter visibility. We set the EOM voltage to the maximum visibility point and maximize it using FPC 1 (i.e.\ we align the polarization of the signal) while compensating with FPC 2 for the lock pulses. A typical interference pattern is shown in Fig.~\ref{Fig:S5}d. We report an average interferometer visibility of $V_{\text{int}}\approx\SI{94}{\%}$ during the whole experiment.

\subsubsection{Phase calibration}
\label{SI:phase_calibration}
\begin{figure*}[h]
	\includegraphics[width = 1\linewidth]{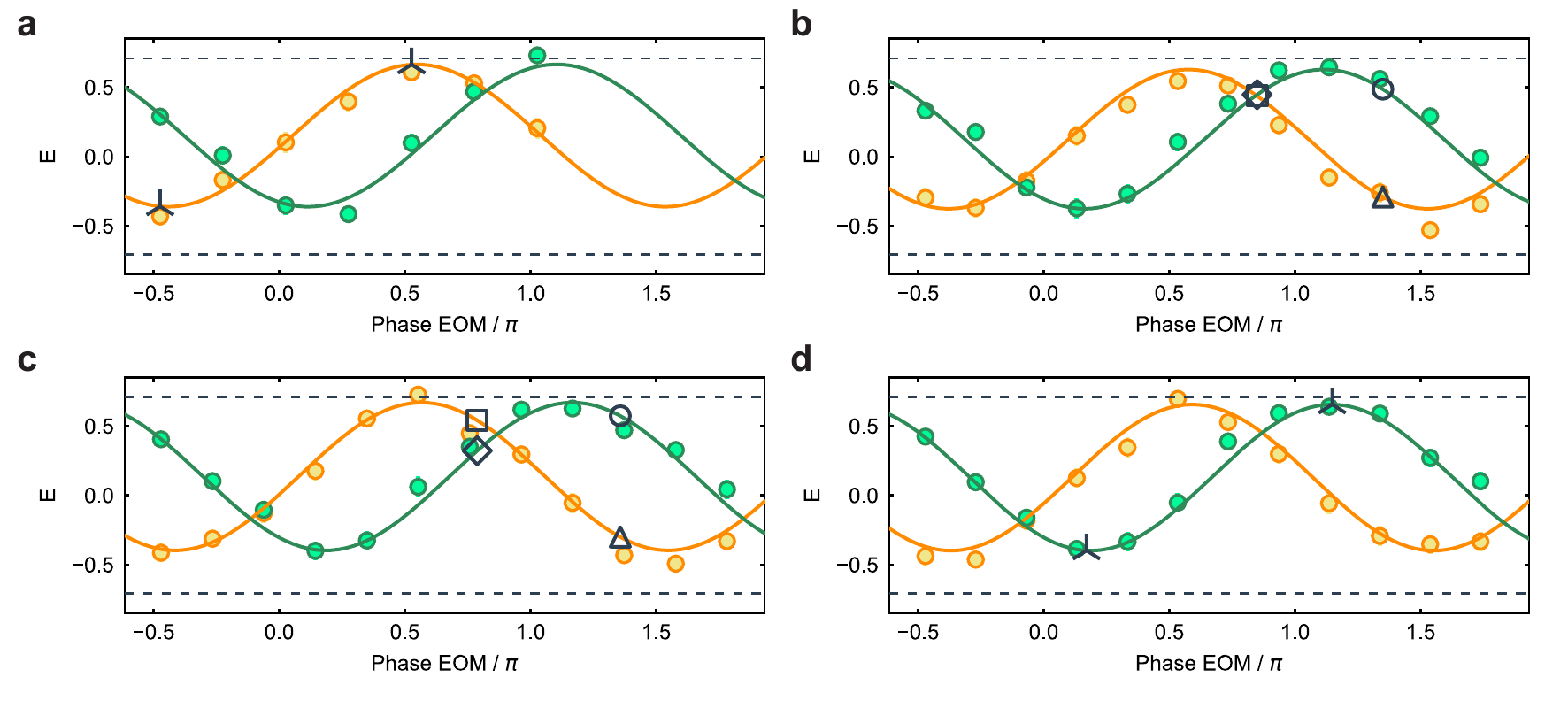}
	\caption{ a) Correlation coefficients $E$, sweeping $\phase{w}$ and for $\phase{r}=0$ (in orange) and $\phase{r}=\pi /2 $ (in green), for higher pulse energies with respect to the main text. The black markers (tri) are at the chosen phases for the additional measure of $R$. b), c) Same as a) for the calibration of the phases for the Bell test. The chosen points differ from the optimal theoretical points of $\phase{0}\pm\pi/4$ by no more than $\approx\pi/20$. The black markers are square, triangle, diamond and circle for the CHSH point $(\pphase{w}{0}, \pphase{r}{0})$, $(\pphase{w}{1}, \pphase{r}{0})$, $(\pphase{w}{0}, \pphase{r}{1})$ and $(\pphase{w}{1}, \pphase{r}{1})$ respectively. d) As reported in a) for the additional measurement of $R$ with $\phase{r}=\pi /2 $. All errors are one standard deviation and are too small to be seen. The small asymmetry in the value of $E$ around zero is a result of the different filter setup efficiencies (see section~\ref{SI:setup}).}
	\label{Fig:S6}
\end{figure*}

To perform the measurements in the main text we need to first accurately calibrate the phase setting. This is done by performing a measurement of $E$ while sweeping $\phase{w}$ and for two settings of $\phase{r}$. Here we use higher pulse energies with respect to the actual measurement reported in the main text, such that the scattering probability increases at the expense of having a lower value of $E$. We use \SI{90}{fJ} (\SI{225}{fJ})) for the write (read) pulses, which gives $p_{\text{w}} = 0.6\%$, $p_{\text{r}} = 1.4\%$. With these settings we obtain more than 200 events per point in about 30 minutes. Fig.~\ref{Fig:S6} shows several such calibration measurements for:

\begin{itemize}
	\item the additional measurement of the $R$ value (Fig.~\ref{Fig:S6}a),
	\item the two runs of integration for the Bell test (Fig.~\ref{Fig:S6}b, c)
	\item and for a final measure of $R$ with $\phase{r}=\pi/2$ (and to check that the phase difference during the second run of integration for the Bell test is small, Fig.~\ref{Fig:S6}d).
\end{itemize}

The values in orange are for $\phase{r}=0$, while $\phase{r}\approx \pi/2$ is shown in green. A small deviation from the desired $\pi /2$ phase difference between the orange and green curves can be seen and the exact values are $\phase{r} =\pi/1.8, \pi/1.9, \pi/1.7, \pi/1.8$, for Fig.~\ref{Fig:S6}a,b,c,d respectively. For the fits of the two datasets we use a sinusoidal function, which serves as a guide to the eye and to numerically calculate the optimal phase points. For the additional measure of $R$ we simply use the phases where $E$ is maximum and minimum (tri markers, Fig.~\ref{Fig:S6}a and d). For the Bell test, instead, we use the fit to numerically calculate the expected $S$ value and choose the phases where the expected $S$ is maximum (square, triangle, diamond and circle markers for the CHSH point $(\pphase{w}{0}, \pphase{r}{0})$, $(\pphase{w}{1}, \pphase{r}{0})$, $(\pphase{w}{0}, \pphase{r}{1})$ and $(\pphase{w}{1}, \pphase{r}{1})$ respectively). In doing so, the experimentally obtained values differ slightly from the theoretical optimal point of $\phase{0}-\pi/4$ and $\phase{0}+\pi/4$ by a small margin $\varepsilon$. The value for $\varepsilon$ for the data in Fig.~\ref{Fig:S6}b (c) is $\approx\pi/30$ and $\approx\pi/20$ ($\approx-\pi/40$ and $\approx\pi/20$), respectively. The phase offset $\phase{\text{off}}$ is calibrated using the maximum and minimum point of $E$. We choose this particular calibration method to compensate for eventual drifts in the phase offset ($\phase{\text{off}}$), as well as small inaccuracies of phase difference for two sets of measurements with different $\phase{r}$. Using light to lock the interferometer at a different frequency and from a different path from that of the signal, gives rise to a (fixed) phase offset $\phase{\text{off}}$ in the entangled state (see section~\ref{SI:setup}). Note that without an external reference PBS a relative change in the polarization between lock pulses and signal pulses will cause a change in the phase shift $\phase{\text{off}}$. However, in our case, the relative change in $\phase{0}$ (equivalently for $\phase{\text{off}}$) is less than $\pi / 50$ in all four measurements.

To further avoid that phase drifts affect only parts of the datasets, we integrate for one hour at each phase point at a time. We then cycle the chosen phases 4 times for phononic entanglement data (and 12 for the longer integration points), and 16 times for the Bell test.

\subsubsection{Effect of non-constant FSR}
\label{SI:FSR_effect}

\begin{figure*}[h]
	\includegraphics[width = 1\linewidth]{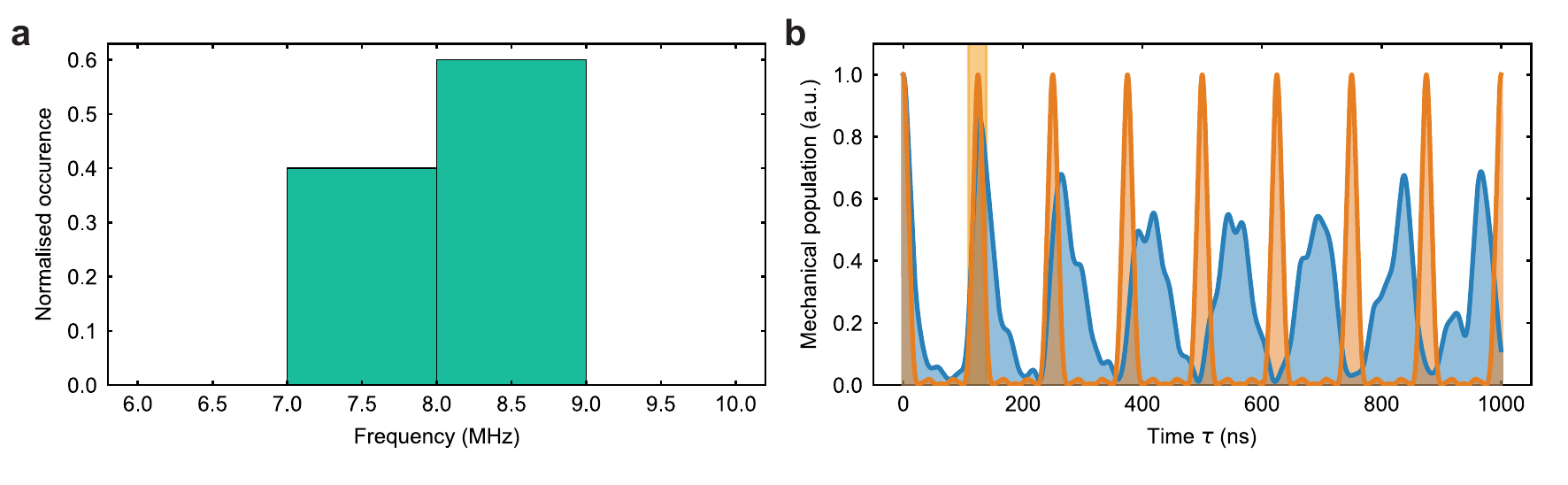}
	\caption{a) Normalized histogram of the FSR between the mechanical peaks shown in Fig.~\ref{Fig:2}c of the main text with mean value of~\SI{8.3}{MHz} and standard deviation~\SI{0.8}{MHz}. b) The numerically calculated time-domain mechanical population (normalized) for the device spectrum (blue) and ideal spectrum with constant FSR (orange). The shaded area is the same as in Fig.~\ref{Eq:2}c.
	}
	\label{Fig:S7}
\end{figure*}

The small dispersion in the waveguide causes a non-constant FSR between the mechanical peaks (see Fig.~\ref{Fig:2}b). In Fig.~\ref{Fig:S7}a we report the histogram of the FSR between the mechanical modes. By using the frequency and amplitude of each mechanical peak in the spectrum, we numerically simulate the time-domain behavior of the mechanical system and compare it with the simulation of the ideal case (i.e.\ with perfectly constant FSR)~\cite{Zivari2022}. As can be seen in Fig.~\ref{Fig:S7}b (blue graph), the mechanical packet is broadened and dimmer after several round-trips due to dispersion of the waveguide compared to the ideal case (orange graph). In this calculation the energy decay of the phonons has not been considered, tracing out any mechanical dissipation, and thus only taking the effect of the dispersion and non-constant FSR into account.

\end{document}